\documentclass[12pt]{article}
\input xy
\xyoption{all}
\input epsf.tex
\usepackage{pstricks}

\usepackage{graphicx}
\usepackage{mathrsfs}
\usepackage{cancel}
\usepackage{amssymb,amsmath}

\def\harr#1#2{\smash{\mathop{\hbox to .3in{\rightarrowfill}}
 \limits^{\scriptstyle#1}_{\scriptstyle#2}}}

\def\s2{\frac{1}{\sqrt2}}

\def\be{\begin{equation}}
\def\ee{\end{equation}}
\def\beqa{\begin{eqnarray}}
\def\eeqa{\end{eqnarray}}

\def\Dsl{\,\raise.15ex\hbox{/}\mkern-13.5mu D} 
\def\d3{d^3}

\begin{document}

\begin{center}
\Large{\bf Rarita-Schwinger Quantum Free Field Via Deformation Quantization}\\
\vspace{1cm}

\large B. Carballo P\'erez\footnote{e-mail address: {\tt
brendacp@fis.cinvestav.mx}}, H. Garc\'{\i}a-Compe\'an\footnote{e-mail address: {\tt
compean@fis.cinvestav.mx}}
\\
[2mm]
{\small \em Departamento de F\'{\i}sica, Centro de
Investigaci\'on y de
Estudios Avanzados del IPN}\\
{\small\em P.O. Box 14-740, 07000 M\'exico D.F., M\'exico}\\

\vspace*{2cm}
\small{\bf Abstract}
\end{center}

\begin{center}
\begin{minipage}[h]{14.0cm} {Rarita-Schwinger (RS) quantum free field is reexamined in the context of deformation quantization. It is found out that the subsidiary condition does not introduce any change either in the Wigner function or in other aspects of the deformation quantization formalism, in relation to the Dirac field case. This happens because the vector structure of the RS field imposes constraints on the space of wave function solutions and not on the operator structure. The RS propagator was also calculated within this formalism.}
\end{minipage}
\end{center}

\bigskip
\bigskip

\date{\today}

\vspace{3cm}

\newpage

\section{Introduction}
\hspace{0.5cm} Deformation quantization (DQ) is nowadays a mathematically well structured and established procedure. It can be regarded as a deformation of the algebraic structure of the ring of functions on the phase space. These functions turn into formal series containing a formal parameter \cite{wwmoriginal,bayen} (for some recent reviews, see \cite{reviews}). However, there are an amount of cases where the formal parameter is a complex (or real) number and it describes real quantization of physical systems \cite{bayen}. Most of the studies of deformation quantization come from the mathematical point of view \cite{wilde,fedosov,k,witten}. One of the standard examples of DQ is the Weyl-Wigner-Groenewold-Moyal (WWGM) formalism \cite{wwmoriginal}, which is valid when the phase space is flat and euclidean. Also it requires from the Schr\"odinger representation of field theory. This is not a standard representation in quantum field theory and it has been useful when one try to gain some intuition from energy eigenvalues in a determined field theory. This strategy has been particularly useful in the deformation quantization of various fields.

Regarding quantum field theory, just a few examples have been considered in the literature for the case of scalar fields \cite{escalar1,escalar2,escalar3,escalar4,escalar5}. The free electromagnetic field and the Dirac field were discussed in \cite{campos,imeldadirac} respectively. Also some works have been done in relation to the gravitational field \cite{antonsen} and bosonic string theory \cite{cuerdas}.  Thus, it is natural to wonder about fields with higher spins. In particular, in the present paper we describe the deformation quantization for the Rarita-Schwinger (RS) free field. This field is of great importance as it describes the gravitino arising in supergravity and superstrings.

In order to do this we briefly review this free field over
Minkowski spacetime $M^{3+1}=\mathbb{R}^3 \times \mathbb{R}$ with
the signature $(-,-,-,+)$, where $x=(\vec{x},t) \in M$. The RS field, $\psi^\mu_\alpha(x)$, where $\alpha$ is an spinorial index ($\alpha=1,\cdots,4$) and $\mu$ is a spacetime vector index ($\mu = 0,1,2,3$), transforms under the Lorentz group as
$({1 \over 2},{1 \over 2}) \otimes \big[({1 \over 2},0) \oplus (0,{1 \over 2})\big] =
({1 \over 2},0) \oplus (0,{1 \over 2}) \oplus (1,{1 \over 2}) \oplus  ({1 \over 2},1)$.

Thus, the field $\psi^\mu({x})$ simultaneously fulfills the Dirac
equation
\begin{equation}
\big(i\not \! \partial - {mc\over\hbar} \big) \psi^\mu (\vec{x},t) =
0
\label{diraceqn}
\end{equation}
and the subsidiary condition
\begin{equation}
\gamma^\mu \psi_\mu =0;
\label{subsidiary}
\end{equation}
where, as usual,
$\gamma^i = \beta \alpha^i$, $\gamma^0 = \beta,$ with $i=1,2,3$.
Note that we use the Weyl (or chiral) representation of the Dirac
matrices $\gamma^\mu$:
\begin{equation}
\begin{array}{ccc}
  \gamma^0 = \left(%
\begin{array}{cc}
  1 & 0 \\
  0 & -1 \\
\end{array}%
\right), & \hspace{0.5cm}  \gamma^j = \left(%
\begin{array}{cc}
  0 & \sigma_j \\
  -\sigma_j & 0 \\
\end{array}%
\right), &\hspace{0.5cm} \\
\end{array}
\end{equation}
where $\sigma_j$ are the Pauli's matrices and $j=1,2,3$. The Dirac matrices have
the following properties: $ {\gamma^j}^\dag = -\gamma^j,$
${\gamma^0}^\dag=\gamma^0,$ $ \gamma^\mu \gamma^\nu + \gamma^\nu
\gamma^\mu = 2\eta^{\mu\nu}.$

 The paper is organized as follows: in Sec. 2 we implement the DQ procedure to quantize the free RS field and finally, in Sec. 3,
  we give our final remarks.

\section{Deformation Quantization of the Rarita-Schwinger Field}

\hspace{0.5cm} If we want to apply the WWGM formalism to the RS free field, one needs to take into account not only the Dirac equation (\ref{diraceqn}), but also the subsidiary condition (\ref{subsidiary}).

It was shown in \cite{imeldadirac} that, for the system described by the Dirac equation (\ref{diraceqn}), the time independent Schr\"odinger equation can be written as

\begin{equation}
\sum_{r=1}^4 \int d^3p \varepsilon_r E_{\vec{p}} {\delta \over
\delta {\bf b}(\vec{p},r)} {\bf b}(\vec{p},r)\Phi[{\bf b}] = E
\Phi[{\bf b}], \label{DiracSch}
\end{equation}
where the wave functional is defined by
$\Phi[{\bf b}] = \langle {\bf b} | \Phi \rangle$, $E_{\vec{p}} =\sqrt{c^2\vec{p}^2 + m^2c^4}$, $\varepsilon_r=1$
for $r=1,2$ and $\varepsilon_r=-1$ for $r=3,4$, ${\bf b}(\vec{p},r)$ is the eigenvalue of $\widehat{\bf
b}(\vec{p},r)$ and it is related to the standard variables $b_r$ and $d_r$ (associated with $\psi^\mu (\vec{x},t)$), as follows
$$
{\bf b}(\vec{p},t,r)= (2 \pi\hbar)^{-3/2}\sqrt{ mc^2
\over E_{\vec{p}}} b_r(\vec{p},t), \hspace{1.0cm} {\rm with\ }
r=1,2,
$$
and
$$
{\bf b}(\vec{p},t,r)= (2 \pi\hbar)^{-3/2}\sqrt{ mc^2 \over
E_{\vec{p}}} d^*_{r-2}(-\vec{p},t), \hspace{1.0cm} {\rm with\ }
r=3,4.
$$
Remember that ${\bf b}^*$ is determined by the complex conjugate of these equations.

Similarly, we can write the subsidiary condition in the form
\begin{equation}
\sum_{r=1}^4 \int {d^3p \over (2
\pi\hbar)^{3/2}} \sqrt{mc^2 \over E_{\vec{p}}}\widehat{\bf
b}(\vec{p},r)\gamma^\mu w_\mu(\vec{p},r) \exp \big(i\vec{p} \cdot \vec{x} / \hbar
\big)\Phi[{\bf b}]=0,
\label{subSch}
\end{equation}
where $w_\mu(\vec{p},r)$ is a solution of the RS equation.

Writing down the equations (\ref{DiracSch}) and (\ref{subSch}) in the WMGM formalism, we have:

\begin{equation}
\sum_{r=1}^4 \int d^3p \varepsilon_r E_{\vec{p}}  {\bf b}^*(\vec{p},r){\bf b}(\vec{p},r)\star{\rho_{W}}[{\bf b}^*,{\bf b}] = E {\rho_{W}}[{\bf b}^*,{\bf b}],
\label{DiracWigner}
\end{equation}
and
\begin{equation}
\sum_{r=1}^4 \int {d^3p \over (2
\pi\hbar)^{3/2}} \sqrt{mc^2 \over E_{\vec{p}}}{\bf
b}(\vec{p},r)\gamma^\mu w_\mu(\vec{p},r) \exp \big(i\vec{p} \cdot \vec{x} / \hbar
\big)\star{\rho_{W}}[{\bf b}^*,{\bf b}]=0.
\label{subWigner}
\end{equation}

For the ground state, the Dirac equation (\ref{DiracWigner}) is equivalent to
\begin{equation}
{\bf b}(\vec{p},r=1,2)  \star {\rho_{W}}_{0}[{\bf b}^*,{\bf b}]= 0, \quad {\bf b}^*(\vec{p},r=3,4)  \star {\rho_{W}}_{0}[{\bf b}^*,{\bf b}]= 0,
\label{gee}
\end{equation}
whose solution (including the normalization factor)
is given by (see ref. \cite{imeldadirac}):
\begin{equation}
{\rho_{W}}_{0}[{\bf b}^*,{\bf b}]  = 2^{-\infty} \exp \bigg\{- {2}
\int {d^3p } \sum_{r=1}^4 \varepsilon_r{\bf b}^*(\vec{p},r) {\bf
b}(\vec{p},r) \bigg\}.
\label{wfgsb}
\end{equation}

Here we took into account the expression for the Moyal $\star$-product
\begin{equation}
\big(F \star G\big)[{\bf b}^*,{\bf b}] = F[{\bf b}^*,{\bf
b}] \exp\bigg({i\hbar\over 2} \buildrel{\leftrightarrow} \over
{\cal P}_{RS}\bigg) G[{\bf b}^*,{\bf b}], \label{fmoyalp}
\end{equation}
where $F[{\bf b}^*,{\bf b}]$
and $G[{\bf b}^*,{\bf b}]$ are functionals over the RS phase
space defined by: ${\cal Z}_{RS} = \{(
{\pi_{\psi}}^\mu_{\alpha}(\vec{x}),{\psi}^\mu_{\alpha}(\vec{x}))_{\vec{x}\in
\mathbb{R}^3}\} = \{(i \hbar {\bf b}^*(\vec{p},r), ({\bf
b}(\vec{p},r))_{r=1,\cdots,4}\}$ and
\begin{equation}
\buildrel{\leftrightarrow}\over {\cal P}_{RS} := -{i \over \hbar}
\sum_{r=1}^{4}\int {d^3p }\bigg( {{\buildrel{\leftarrow}\over
{\delta}}\over \delta {\bf b}(\vec{p},r)}
{{\buildrel{\rightarrow}\over {\delta}}\over \delta {\bf
b}^*(\vec{p},r)} + {{\buildrel{\leftarrow}\over {\delta}}\over
\delta {\bf b}^*(\vec{p},r)} {{\buildrel{\rightarrow}\over
{\delta}}\over \delta {\bf b}(\vec{p},r)}  \bigg).
\label{dpoisson}
\end{equation}

As $\gamma^\mu w_\mu(\vec{p},r)=0$, equation (\ref{subWigner}) does not impose new restrictions on ${\bf b}(\vec{p},r)$ or ${\bf b}^*(\vec{p},r)$. That is the reason why there are no changes in the Wigner function due to the consideration of the subsidiary condition.

The Winger functionals for excited states can be constructed, as in \cite{imeldadirac}, in the following form
$$
{\rho_{W}}[{\bf b}^*,{\bf b}] ={\bf b}^{*}(\vec{p}_1,r_1) \cdots
{\bf b}^{*} (\vec{p}_n,r_n) {\bf b}(-\vec{p}_{1'},r_{1'})\cdots
{\bf b}(-\vec{p}_{n'},r_{n'}) \star {\rho_{W}}_{0}
$$
\begin{equation}
\star {\bf b}^{*}(-\vec{p}_{n'},r_{n'})\cdots {\bf
b}^{*}(-\vec{p}_{1'},r_{1'})  {\bf b}(\vec{p}_n,r_n) \cdots {\bf
b}(\vec{p}_1,r_1).
\label{fexcststesII}
\end{equation}
From this last formula one can find the Wigner functional for any
excited state, which will be also identical to the Dirac case.

The other aspects of the WWGM, such as: the Stratonovich-Weyl quantizer and normal ordering are also the same as for the Dirac case. However, the propagator depending on the wave functions that contains the information about the solutions of positive and negative energy of the RS equation, would be different. In the next section we compute it using the WWGM formalism.

\subsection{Rarita-Schwinger Propagator}

\hspace{0.5cm} In order to compute the propagator of the RS field we need to
find
\begin{equation}
 iS_F({x}-{y}) = \langle
0|\widehat{\psi^{\mu}}_{\alpha}({x}) \widehat{\overline{\psi^{\mu}}}_{\beta}({y})|0\rangle \cdot
\theta(t-t') - \langle 0|
\widehat{\overline{\psi^{\mu}}}_{\beta}({y}) \widehat{{\psi}^{\mu}}_{\alpha}({x})|0\rangle \cdot
\theta(t'-t). \label{propa}
\end{equation}

So we first compute the quantities $\langle 0|\widehat{\psi}^{\mu}_{\alpha}({x})
\widehat{\overline{\psi^{\mu}}}_{\beta}({y})|0\rangle $ and $\langle 0|
\widehat{\overline{\psi^{\mu}}}_{\beta}({y})\psi^{\mu}_{\alpha}({x})|0\rangle$. In terms
of deformation quantization these expectation values are given by (compare with
\cite{imeldadirac})

\begin{equation}
\langle 0 |\widehat{\psi}^{\mu}_{\alpha}({x}) \widehat{\overline{\psi^{\mu}}}_{\beta}({y})| 0
\rangle =
 {\int \prod d{\bf b}^* d {\bf b} \ \psi^{\mu}_{\alpha}({x})
 \star
\overline{\psi^{\mu}}_{\beta}({y}) \ \rho_{W_0} [{\bf b}^*,{\bf b}]
\over \int \prod d{\bf b}^* d {\bf b} \ \rho_{W_0} [{\bf b}^*,{\bf
b}]}, \label{promedio}
\end{equation}
and the analogous formula for the second expectation value.

Carrying out the corresponding integrations and making use of the
following relations (see ref. \cite{Lurie}):
\begin{eqnarray}
\sum_{r=1}^2 w^{\mu}_{\alpha}(\vec{p},r) \overline{w}^{\mu}_{\beta}(\vec{p},r)
= {(\not \! \! p + mc)\over 2mc}[\delta^{\alpha\beta}-\frac{1}{3}\gamma^\alpha\gamma^\beta-\frac{1}{3mc}(\gamma^\alpha p^\beta - \gamma^\beta p^\alpha)-\frac{2}{3m^2 c^2}p^\alpha p^\beta],\\
\sum_{r=3}^4 \overline{w}^{\mu}_{\alpha}(\vec{p},r) w^{\mu}_{\beta}(\vec{p},r)
= {(\not \! \! p - mc)\over 2mc}[\delta^{\alpha\beta}-\frac{1}{3}\gamma^\alpha\gamma^\beta+\frac{1}{3mc}(\gamma^\alpha p^\beta - \gamma^\beta p^\alpha)-\frac{2}{3m^2 c^2}p^\alpha p^\beta],
\end{eqnarray}
after straightforward calculations we arrive at the following results
\begin{eqnarray}
\langle 0 |\widehat{\psi}^{\mu}_{\alpha}({x})
\widehat{\overline{\psi}}^{\mu}_{\beta}({y})| 0 \rangle =  \int {d^3p
\over (2 \pi\hbar)^3}{c(\not \! \! p + mc)\over
2E_{\vec{p}}}[\delta^{\alpha\beta}-\frac{1}{3}\gamma^\alpha\gamma^\beta-\frac{1}{3mc}(\gamma^\alpha p^\beta - \gamma^\beta p^\alpha)\nonumber\\
-\frac{2}{3m^2 c^2}p^\alpha p^\beta] \exp \big(-i p\cdot(x-y)/\hbar\big);
\label{cienvcinco}
\end{eqnarray}
and
\begin{eqnarray}
 \langle 0|
\widehat{\overline{\psi}}^{\mu}_{\beta}({y}) \widehat{\psi}^{\mu}_{\alpha}({x})|0\rangle =
  \int {d^3p \over (2 \pi\hbar)^3}{ c(\not \! \! p -
mc)\over 2E_{\vec{p}}}[\delta^{\alpha\beta}-\frac{1}{3}\gamma^\alpha\gamma^\beta+\frac{1}{3mc}(\gamma^\alpha p^\beta - \gamma^\beta p^\alpha)\nonumber\\
-\frac{2}{3m^2 c^2}p^\alpha p^\beta] \exp \big(i
p\cdot(x-y)/\hbar\big),
\label{otro}
\end{eqnarray}
where $p \cdot x= \eta^{\mu \nu} p_\mu x_\nu.$ The above formulas reproduce exactly the propagator of the RS field.


\section{Concluding Remarks}

\hspace{0.5cm}In this paper we have extended the deformation quantization
program via the WWGM formalism from Dirac to RS free field.
We found out that the subsidiary equation does not introduce any change in the Wigner function of the Dirac field. For that reason, our main result is that the subsidiary condition does not affect the quantization of this field. Consequently, the quantum states are the same as those for the Dirac case. The difference appears only at the level of the space of the wave functions which are solutions of the equation of motion (\ref{diraceqn}) and the subsidiary condition (\ref{subsidiary}). Then, although the quantum states are the same, the propagator and also the scattering amplitudes, both determined by the dynamics of the theory, are very different.

It is interesting to note that something similar happens with the CPT group of the Rarita-Schwinger field in relation to the CPT group of the Dirac field: both fermionic fields have the same CPT group \cite{CPT RS}.

In the case of deformation quantization as well as in the case of the CPT group, there is no a priori reason to think that the results for the Dirac field and for the Rarita-Schwinger field must be coinciding. Both results show that these two fields share similar properties despite their different nature.

Making a comparison for the spin 3/2 field between the procedures of deformation quantization and canonical quantization, we can find out that the latter becomes rather awkward due to the difficulty of isolating the independent dynamical degrees of freedom \cite{Lurie}; while in the first case this problem is avoided because DQ does not distinguish between Dirac and Rarita-Schwinger fields.

\vskip 1truecm
\centerline{\bf Acknowledgments}
The research of B.C.P. is supported by a CLAF-ICyTDF postdoctoral fellowship. The research of H.G.-C. is supported in part by a CONACyT grant 128761.

\bibliography{octaviostrings}

\begin{thebibliography}{99}


\bibitem{wwmoriginal} H. Weyl, {\it Group Theory and Quantum Mechanics}, (Dover, New York, 1931); E.P. Wigner, Phys. Rev. {\bf 40}, 749 (1932); A. Groenewold, Physica {\bf 12} (1946) 405-460; J.E. Moyal, Proc. Camb. Phil. Soc. {\bf 45}, 99 (1949).

\bibitem{bayen} F. Bayen, M. Flato, C. Fronsdal, A. Lichnerowicz and D. Sternheimer, Ann. Phys. {\bf 111}, 61 (1978); Ann. Phys. {\bf 111}, 111 (1978).

\bibitem{reviews} C.K. Zachos, {\it Deformation Quantization: Quantum Mechanics Lives and Works in Phase Space}, Int. J. Mod. Phys. A {\bf 17} (2002) 297, hep-th/0110114; A.C. Hirshfeld and P. Henselder, {\it Deformation Quantization in the Teaching for Quantum Mechanics}, Am. J. Phys. {\bf 70} (2002) 537; G. Dito and D.Sternheimer, {\it Deformation Quantization: Genesis, Developments and Metamorphoses,  Deformation Quantization} (Strasbourg 2001) Lect. Math. Theor. Phys. {\bf 1} Ed. de Gruyter, Berlin, IRMA (2002) pp. 9-54.

\bibitem{wilde} M. De Wilde and P.B.A. Lecomte, Lett. Math. Phys. {\bf 7} (1983) 487; H. Omori, Y. Maeda and A. Yoshioka, Adv. Math. {\bf 85}, 224 (1991).

\bibitem{fedosov} B. Fedosov, J. {\it Diff. Geom.} {\bf 40}, 213 (1994); {\it Deformation Quantization and Index Theory} (Akademie Verlag, Berlin, 1996).

\bibitem{k} M. Kontsevich, {\it Deformation Quantization of Poisson Manifolds I}. [q-alg/9709040]; Lett. Math. Phys. {\bf 48}, 35 (1999).

\bibitem{witten}
  S.~Gukov and E.~Witten,
  {\it Branes and Quantization}, arXiv:0809.0305 [hep-th].

\bibitem{escalar1} G. Dito,  {\it Star Product Approach to Quantum Field Theory: The Free Scalar Field}, Lett. Math. Phys., {\bf 20}, 125 (1990).

\bibitem{escalar2} G. Dito,  {\it Star Products and Nonstandard Quantization for Klein-Gordon Equation}, J. Math. Phys., {\bf 33}, 791 (1992).

\bibitem{escalar3} G. Dito,  {\it An Example of Cancellation of Infitinies in the Star Quantization of Fields}, Lett. Math. Phys., {\bf 27}, 73 (1993).

\bibitem{escalar4} T. Curtright, D. Fairlie and C. K. Zachos, Phys. Rev. D, {\bf 58}, 025002 (1998).

\bibitem{escalar5} T. Curtright and C. K. Zachos, J. Phys. A, {\bf 32}, 771 (1999).

\bibitem{campos} H. Garc\'{\i}a-Compe\'an, J.F. Pleba\'nski, M. Przanowski and F.J. Turrubiates, Int. J. Mod. Phys. A {\bf 16}, 2533 (2001).

\bibitem{imeldadirac}
  I.~Galaviz, H.~Garcia-Compean, M.~Przanowski and F.~J.~Turrubiates,
  Annals Phys.\  {\bf 323}, 827 (2008)
  [arXiv:hep-th/0703125].

\bibitem{antonsen} F. Antonsen,  Phys. Rev. D, {\bf 56}, 920 (1997); {\it Deformation Quantization of Constrained Systems}, gr-qc/9710021.

\bibitem{cuerdas}
H. Garc\'{\i}a-Compe\'an, J. F. Pleba\'nski, M. Przanowski and F.J.Turrubiates, J. Phys. A: Math. Gen., {\bf 33}, 7935 (2000).

\bibitem{Lurie} D. Luri\'e, \emph{Particles and Fields} (Interscience Publishers), pp. 44-51 (1968).

\bibitem{CPT RS} B. Carballo P\'erez and M. Socolovsky, {\it The CPT group of the spin-3/2 field} ,  arXiv: hep-th/1001.0751v2 (2010).




\end{thebibliography}
\addcontentsline{toc}{section}{Bibliography}
\bibliographystyle{TitleAndArxiv}


\end{document}